# Does the Coulomb potential have an algebraic origin?

Anzor Khelashvili

*Institute of High Energy Physics, Iv. Javakhishvili Tbilisi State University, University Str. 9, 0109, Tbilisi, Georgia. E-mail: anzor.khelashvili@tsu.ge*

**Abstract:** It is shown that in case of central potentials, both the fourth component of Lorentz vector as well as Lorentz scalar in the Dirac Hamiltonian, owing to the conserved Dirac spin-orbital matrix, there arises Witten's N=2 superalgebra. The generators of this algebra are constucted and their commutativity with Dirac Hamiltonian is studied. Under the requirement of invariance relative to this superalgebra it follows that only Coulomb-like potential obeys the corresponding constraints. This fact allows us to suppose the Witten's superalgebra as an alternative source for emerging of the Coulomb potential. As a byproduct, we obtain energy spectrum without solving the Dirac equation, i.e. by pure algebraically.

## I.     Introduction

Long-range $1/r$ like potentials play a fundamental role in physics. Their origin is usually traced back to the existence of massless particles as photons or gravitons related to fundamental properties of quantum field theories as gauge invariance. In this article it is argued that, in principle, $1/r$ potential might also occur in relativistic quantum mechanics, i.e. in Dirac's theory, where the simplest Witten's superalgebra appears in very natural way.

Particle Physicists are pinning their latest hope on the Supersymmetry (SUSY). It is a very rich theory, but has no motivation. Moreover, this theory does not fixes its scale, and in general, there is no indication where is the threshold of this symmetry. Nevertheless, humanity takes the risk and is going to build new colliders and accelerators by fantastically rising energies. Some of the projects are now considered seriously. The truth is that the theory is at fault, but SUSY is very attractive and opens many new interesting possibilities, such as inclusion of gravity into the scheme. I do not consider such delicate problems here, but I have restricted myself by the frame of SUSY quantum mechanics, developed by Witten early in eighties of previous century.

I consider the Dirac Hamiltonian in 3-dimensions. This Hamiltonian is the first relativistic problem, where the spin appears and the simplest Witten's superalgebra is realized.

## II. Witten's superalgebra in the Dirac Hamiltonian

Let us consider the Dirac Hamiltonian in the central-potential field

$$H_D = \boldsymbol{\alpha} \cdot \boldsymbol{p} + \beta m + V(r) \qquad (1)$$

We are dealing with 3-dimensional physical problem (the Dirac's representation will be used for matrices). It is well known that the orbital momentum operator $\boldsymbol{l} = \boldsymbol{r} \times \boldsymbol{p}$ and the spin matrix $\Sigma = diag(\sigma, \sigma) = \begin{pmatrix} \sigma, & 0 \\ 0, & \sigma \end{pmatrix}$ do not separately commute with the Hamiltonian,

$$[l_i, H] = [l_i, \boldsymbol{\alpha} \cdot \boldsymbol{p}] = i\varepsilon_{ijk}\alpha_j p_k, \qquad \frac{1}{2}[\Sigma_i, H] = \frac{1}{2}[\Sigma_i, \boldsymbol{\alpha} \cdot \boldsymbol{p}] = -i\varepsilon_{ijk}\alpha_j p_k \qquad (2)$$

But the Total momentum operator $\boldsymbol{J} = \boldsymbol{l} + \frac{1}{2}\Sigma$ commutes with it. Therefore the total momentum is conserved.

Moreover, as is well known, there is one more operator, the Dirac's spin-orbital matrix

$$K = \beta(\Sigma \cdot \boldsymbol{l} + 1), \qquad (3)$$

That commutes with the Dirac Hamiltonian in arbitrary central-symmetry field.

Remarkably enough, less attention was devoted to this operator in the classical textbooks. May be, because it is not related to some known transformations. We'll see below that the rather interesting results can be derived by this matrix.

It is evident that this matrix commutes with the beta matrix and anticomputer with $\gamma^5$:

$$[K, \beta] = 0; \qquad \{K, \gamma^5\} = 0 \qquad (4)$$

The question is: are there another operators anticommuting with $K$?

If there is a such operator, let call it $Q$, then any operator of kind $\tilde{Q} = i\dfrac{QK}{\sqrt{K^2}} = i\dfrac{QK}{|\kappa|}$, where $\kappa$ is the eigenvalue of $K$, will also be anticommuting with $K$. Indeed, It is easy to verify, that

$$\{\tilde{Q}, K\} = 0, \quad \text{provided} \quad \{Q, K\} = 0. \qquad (5)$$

What is more, $\quad \tilde{Q}^2 = Q^2.$ \hfill (6)

Let us introduce the notation, for convenience, $Q = Q_1$ and $\tilde{Q} = Q_2$. Then the above relations may be written in the following compact form

$$\begin{aligned}\{Q_i, Q_j\} &= 2\delta_{ij}\,h; & (i, j = 1, 2) \\ [h, Q_k] &= 0; & (k = 1, 2) \\ h &\equiv Q_i^2 \end{aligned} \qquad (7)$$

We see that the so called Witten's N=2 superalgebra [1] is derived almost automatically, where the generators are $Q_1, Q_2$, and plays the role of Witten's Hamiltonian.

Hence we have the Dirac Hamiltonian at the one hand and the Witten's superalgebra on the other one.

### III. Symmetry under superalgebra

There one question arises: In what circumstances' may happen that the Dirac Hamiltonian becomes Symmetrical under the mentioned Witten's superalgebra? Or for which potential do we have the commutativity

$$[Q_i, H_D] = 0 \;? \tag{8}$$

For this aim let us use the theorem, proved by us in [2,3]:

### Theorem:

*Let $V$ is a vector with respect of the angular momentum, $l$:*

$$[l_i, V_j] = i\varepsilon_{ijk} V_k$$

*Suppose that this vector is perpendicular to $l$, i.e, $(l \cdot V) = (V \cdot l) = 0$. Then it is easily to verify, that the Dirac matrix $K$ anticommutes with $(\Sigma \cdot V)$, which is scalar under total momentum $J$.*

It is evident that the class of anticommuting operators may be enlarged – any operator like $\hat{O}(\Sigma \cdot V)$, where $\hat{O}$ commutes with $K$, will be also $K$-odd, i.e. anticommutes with $K$.

By using of physical vectors at hand we are able to compose the most general operator that obeys to the condition of above theorem. It is:

$$Q = x_1(\Sigma \cdot \hat{r}) + ix_2 K(\Sigma \cdot p) + ix_3 K\gamma^5 f(r) \tag{9}$$

Here the arbitrary coefficients are chosen such, that the Hermitian operator be obtained. $x_{1,2,3}$ are the real numbers, and $f(r)$ is an arbitrary real scalar function to be determined by symmetry requirement.

If one finds such a $Q$, then we'll find also the second generator $\tilde{Q}$, and, therefore reconstruct all generators of the Witten superalgebra.

Now we can require the symmetry under this algebra or require commutativity of suggested generator with the Dirac's Hamiltonian. For this aim compute the needed commutator and equate it to zero

$$[Q, H] = (\Sigma \cdot \hat{r})\left[x_2 V'(r) - x_3 f'(r)\right] + 2i\beta K\gamma^5 \left[\frac{x_1}{r} - mf(r)x_3\right] = 0 \tag{10}$$

This expression must be zero. Evidently it happens only if the coefficients of diagonal and antidiagonal matrix-elements could be zero. It means:

$$x_2 V'(r) = x_3 f'(r)$$
$$x_3 m f(r) = \frac{x_1}{r}$$
(11)

It follows uniquely very important statement about the potential

$$V(r) = \frac{x_1}{m x_2} \frac{1}{r}$$
(12)

Hence, in the framework of very general considerations we have shown that *the only central potential for which the Dirac Hamiltonian has an additional symmetry - N=2 supersymmetry in the above sense – is a Coulomb potential.*

It follows that the form of the Coulomb potential automatically results from the requirement of symmetry of the Dirac's Hamiltonian with respect to the Witten's N=2 superalgebra, or *this algebra can be considered as a source of Coulomb potential.*

## IV. The Physical meaning of supersymmetry generators

If we take into account the derived restrictions, it follows for the supersymmetry generator

$$Q = x_1 \left\{ (\Sigma \cdot \hat{r}) - \frac{i}{ma} K (\Sigma \cdot p) + \frac{i}{mr} K \gamma^5 \right\}$$
(13)

This expression is odd one and satisfies all the conditions of above theorem. After this results was obtained, we found that it was an alternative form of operators, published by Johnson and Lippmann [4] as an abstract in Physical Review in 1950: **M.H.Johnson, B.A.Lippmann**, "Relativistic Kepler Problem", Phys. Rev. **78**, 329 (1950).

Abstract: *"Besides the usual integrals of motion $\vec{M}$ and $j$ (in Dirac's notation) the relativistic equations for a charge in a Coulomb field admit*

$$A' = \vec{\sigma} \vec{r} r^{-1} - i \left( \frac{\hbar c}{e^2} \right) (mc^2)^{-1} j \rho_1 (H - mc^2 \rho_1)$$

*as another integral of motion. Since A and j anticommute, the pairs with the same |j| are degenerate. Thus the existence of A estabilashed the "accidental" degeneracy with respect to l in the corresponding non-relativistic problem".*

In modern notations this operator looks like the above one if we neglect the irrelevant numerical factor

$$A' = \gamma^5 \left\{ \boldsymbol{\alpha} \cdot \hat{\boldsymbol{r}} - \frac{i}{ma} K \gamma^5 (H - \beta m) \right\} \tag{14}$$

Therefore we have derived the Johnson-Lippmann operator and at the same time demonstrated its commutativity with the Dirac-Coulomb Hamiltonian.

One can show that the following relation takes place:

$$K(\boldsymbol{\Sigma} \cdot \boldsymbol{V}) = -i\beta \left( \boldsymbol{\Sigma}, \frac{1}{2} [\boldsymbol{V} \times \boldsymbol{l} - \boldsymbol{l} \times \boldsymbol{V}] \right) \tag{15}$$

With the help of it our operator reads

$$A' = \boldsymbol{\Sigma} \cdot \left\{ \hat{\boldsymbol{r}} - \frac{i}{2ma} \beta (\boldsymbol{p} \times \boldsymbol{l} - \boldsymbol{l} \times \boldsymbol{p}) \right\} + \frac{i}{mr} K \gamma^5 \tag{16}$$

Consider now its non-relativistic limit, when

$$\beta \to 1, \quad \gamma^5 \to 0 \tag{17}$$

In this limit our operator becomes

$$A' = \boldsymbol{\Sigma} \cdot \boldsymbol{A}, \tag{18}$$

where

$$\boldsymbol{A} = \hat{\boldsymbol{r}} - \frac{i}{2ma} (\boldsymbol{p} \times \boldsymbol{l} - \boldsymbol{l} \times \boldsymbol{p}) \tag{19}$$

It is the known Laplace-Runge-Lentz (LRL)vector for the Coulomb potential – hence our operator is a projection of LRL vector on the spin direction

After that it is clear that one derive the Witten's algebra by the following identification

$$Q_1 = Q, \qquad Q_2 = \tilde{Q} = i \frac{KQ}{\kappa}, \tag{20}$$

In what follows the Witten's N=2 superalgebra

$$\{Q_1, Q_2\} = 0, \qquad Q_1^2 = Q_2^2 = Q^2 \tag{21}$$

Corresponding nilpotent operators can be constructed simply as $Q_\pm = Q_1 \pm iQ_2$. Evidently $Q_\pm^2 = 0$.

It is well-known from Classical Mechanics that the orbit equations follow by squaring the LRL vector.

In our case the square of the obtained generator gives

$$Q^2 = 1 + \left(\frac{K}{a}\right)^2 \left(\frac{H^2}{m^2} - 1\right) \tag{22}$$

All operators entering here commute with each other. Therefore one can replace them by their simultaneous eigenvalues. For the ground state we must have $Q^2 = 0$. It gives automatically the ground state energy

$$\frac{E}{m} = \sqrt{1 - \left(\frac{Z\alpha}{\kappa}\right)^2} = \frac{1}{\sqrt{1 + \left(\frac{Z\alpha}{s}\right)^2}}, \tag{23}$$

$$s \equiv \sqrt{\kappa^2 - (Z\alpha)^2} \tag{24}$$

One can perform the ladder procedure for obtaining other states. It is demonstrated in paper [5] that this procedure reduces to replacement $\sqrt{\kappa^2 - a^2} \rightarrow \sqrt{\kappa^2 - a^2} + n - |\kappa|$, after that one obtains the known Sommerfeld (1916) formula

$$E = m \left\{ 1 + \frac{(Za)^2}{\left(n - |\kappa| + \sqrt{\kappa^2 - (Za)^2}\right)^2} \right\}^{-1/2} \tag{25}$$

It is the very "strange" fact that this formula was obtained by Sommerfeld, in time when the Dirac equation and spin was not jet known. About this peculiarity of the Coulomb potential wrote in its very exhaustive article [6]

It is remarkable that the double degeneracy of Hydrogen spectrum remains, because it depends only on $|\kappa|$. This degeneracy was removed experimentally (famous *Lamb Schift*). Our consideration shows that this degeneracy is connected with obtained conserved operator – as well as the potential shows the deviation from the pure Coulomb form, the Witten's symmetry breaks immediately and the degeneracy disappears.

For example, one loop corrections to photon propagator and photon-electron vertex give the following additional term to Coulomb potential

$$\Delta V_{eff} \approx \frac{4\alpha^2}{3m^2}(\ln\frac{m}{\mu} - \frac{1}{5})\delta^3(\vec{r}) + \frac{\alpha^2}{2\pi m^2 r^3}(\vec{\Sigma}\cdot\vec{l}) \qquad (26)$$

We see, that this contribution does not commute with our Q-operator. Therefore its inclusion into Hamiltonian will break the Witten's symmetry. It means that *this symmetry is responsible for the forbidden of Lamb shift.*

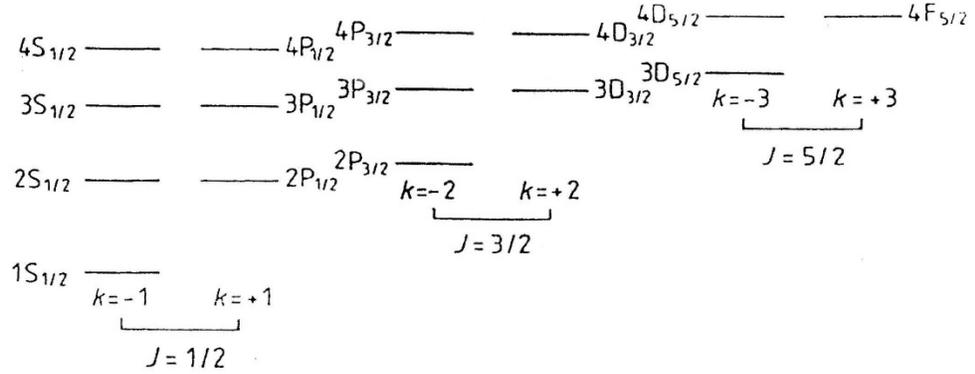

**Fig.1.** Hydrogen atom spectrum according to Sommerfeld formula shows the SUSY structure, as a result of degeneracy $\kappa \to -\kappa$. Only QED removes this degeneracy.

### V.  Generalization to the case of scalar potential

It is interesting that the inclusion of the Lorentz-scalar potential gives the correct generalization of the description method. Let us note that this problem was considered earlier by A.Leviatan [7] on the basis of radial Dirac equations. The author used the radial decomposition of the Dirac equation and studied the problem in terms of Differential operators. After that we considered the same problem in [8], by using the above described method. Now   Let us consider some of critical points: the Dirac Hamiltonian for both scalar and the Lorentz 4th component potential has the form

$$H = \vec{\alpha} \cdot \vec{p} + \beta m + V(r) + \beta S(r) \tag{27}$$

Here $S(r)$ is a Lorentz-scalar potential. This Hamiltonian commutes with $K$-operator, but does not commute with the above derived JL operator.

For deriving a new commuting odd operator, we include new structures, which are permissible by our theorem and consider the following modification

$$X = x_1(\vec{\Sigma} \cdot \hat{\vec{r}}) + x_1'(\vec{\Sigma} \cdot \hat{\vec{r}})H + ix_2 K(\vec{\Sigma} \cdot \vec{p}) + ix_3 K\gamma_5 f_1(r) + ix_3' K\gamma_5 \beta f_2(r) \tag{28}$$

New terms consist factors $\hat{O} = H$ and $\hat{O} = \beta$, they commute with $(\Sigma \cdot V)$ and do not destroy linearity with respect of $\hat{r}$ and $p$ vectors. The previous case follows if we take $S(r) = 0$ and $x_1' = x_3' = 0$. Calculating the commutator with Hamiltonian and equating to zero, we obtain the supplementary conditions:

(1) From antidiagonal structures $(\gamma^5 K, \gamma^5 \beta K)$:

$$\frac{x_1}{r} - x_3(m+S)f_1(r) + \frac{x_1'}{r}V(r) = 0$$

$$\frac{x_1'}{r}(m+S) - (m+S)f_2(r) = 0 \tag{29}$$

**(2)** From diagonal structures ($K(\vec{\Sigma} \cdot \hat{\vec{r}})$, $K\beta(\vec{\Sigma} \cdot \hat{\vec{r}})$, $\beta K(\vec{\Sigma} \cdot \vec{p})$)

$$x_2 V'(r) - x_3 f_1'(r) = 0$$

$$\frac{x_1}{r} - x_2(m+S)V(r) + \frac{x_1'}{r}V(r) = 0$$

$$x_2 S'(r) - x_3' f_2'(r) = 0 \tag{30}$$

$$\frac{x_1'}{r} - x_3' f_2(r) = 0$$

Now from the first and third equations of (30) after integration with zero boundary condition at infinity we obtain the constraints:

$$f_1(r) = \frac{x_2}{x_3} V(r), \qquad f_2(r) = \frac{x_2}{x_3'} S(r) \tag{31}$$

At the same time, the last equation gives

$$f_2(r) = \frac{x_1'}{x_3' r} \tag{32}$$

Therefore

$$S(r) = \frac{x_1'}{x_2 r} \qquad (33)$$

Or, *the scalar potential must be Coulombic one.*

After substituting all of this into the first set of equations, we get:

$$V(r) = \frac{x_1'}{x_2 r} \qquad (34)$$

In conclusion, *both scalar and vector (gauge) potentials must be Coulombic ones, if we want to describe the degeneracy in signs* $\pm \kappa$.

Taking into account all derived solutions into the general expression of $X$ one can construct the corresponding symmetry operator in more compact form [8]

$$X = (\vec{\Sigma} \cdot \hat{\vec{r}})(m a_V + H a_S) - i K \gamma_5 (H - \beta m) \qquad (35)$$

Here the following notations are used:

$$a_V = -\frac{x_1}{x_2 m}, \qquad a_S = -\frac{x_1'}{x_2}$$

and $a_{V,S}$ are the Coulomb coupling constants

$$V(r) = -\frac{a_V}{r}, \qquad S(r) = -\frac{a_S}{r}, \qquad a_{V,S} > 0 \qquad (36)$$

In contrast of A. Leviatan our approach is 3-dimensional, we do not use the Dirac equation (only Hamiltonian and our theorem). It is more systematic, transparent and informative from the point of symmetry, by our opinion.

## VI. Spectrum of the Dirac Equation in general case of both potentials (algebraic derivation)

Now we chose the following identification:

$$Q_1 = X, \qquad Q_2 = i \frac{X K}{|\kappa|} \qquad (37)$$

Then

$$\{Q_1, Q_2\} = 0, \qquad Q_1^2 = Q_2^2 \equiv h \qquad (38)$$

we are faced again with the Witten's hamiltonian.

To explore this algebra, let us define the ground state of supersymmetry as

$$h|0\rangle = X^2|0\rangle = 0 \tag{39}$$

Because the Witten Hamiltonian is a square of the Hermitian operator, its spectrum will be positive definite. Therefore one can account this operator vanishing in vacuum state

$$X|0\rangle = 0 \tag{40}$$

Inserting here the explicit form of operator (33) and obtain the Hamiltonian, it follows

$$H = m[(\vec{\alpha}\cdot\hat{\vec{r}})a_S + iK]^{-1}[iK\beta - a_V(\vec{\alpha}\cdot\hat{\vec{r}})] = \frac{m}{K^2 + a_S^2} N, \tag{41}$$

where

$$N \equiv [(\vec{\alpha}\cdot\hat{\vec{r}})a_S - iK][iK\beta - a_V(\vec{\alpha}\cdot\hat{\vec{r}})] =$$
$$= -a_S a_V + K[K\beta + ia_V(\vec{\alpha}\cdot\hat{\vec{r}})] - ia_S K\beta(\vec{\alpha}\cdot\hat{\vec{r}}) \tag{42}$$

Now by means the Foldy-Wouthuysen like transformation one can diagonalize this Hamiltonian. Owing to the fact that in the last equation the second and third terms do not commute with each other, we need several (at least, two) such transformations.

Let us chose firstly

$$\exp(iS_1) = \exp\left(-\frac{1}{2}\beta(\vec{\alpha}\cdot\hat{\vec{r}})w_1\right), \quad thw_1 = -\frac{a_V}{K} \tag{43}$$

It follows after transforming

$$N \rightarrow N' = -a_S a_V + \beta\sqrt{K^2 - a_V^2} - ia_S K\beta(\boldsymbol{\alpha}\cdot\hat{\boldsymbol{r}}) \tag{44}$$

Now perform the second transformation

$$N'' = \exp(iS_2)N'\exp(-iS_2), \tag{45}$$

where

$$S_2 = -\frac{1}{2}(\vec{\alpha}\cdot\hat{\vec{r}})w_2 \quad \text{and} \quad tgw_2 = \frac{a_S}{\sqrt{K^2 - a_V^2}} \tag{46}$$

Thus,

$$N'' = -a_S a_V + K\beta\sqrt{K^2 - a_V^2 + a_S^2} \tag{47}$$

Finally

$$H = \frac{m}{K^2 + a_S^2}\left\{-a_S a_V + K\sqrt{K^2 - a_V^2 + a_S^2}\beta\right\}, \tag{48}$$

and we find expression for the ground state energy

$$E_0 = \frac{m}{\kappa^2 + a_S^2}\left\{-a_S a_V \pm \kappa\sqrt{\kappa^2 - a_V^2 + a_S^2}\right\} \tag{49}$$

It is the correct expression for the ground state energy, which coincides to the formula, derived by explicit solution of the Dirac equation [9].

This result is really amazing – derivation of spectrum without solving the equation of motion, the combined Coulomb problem is totally integrable one.

So, only on the basis of algebraic methods we obtained the correct energy spectrum. This fact show how effective are application of symmetry methods in cases, when it is possible.

### VII. Conclusions

Above consideration shows that the Witten's superalgebra emerges nearly in itself because existence of conserved Dirac's spin-orbital matrix. The superalgebra generators, constracted according to definite theorem, commute with the Dirac Hamiltonian only for Coulomb potential both for vector as well as scalar ones.

Therefore, we can conclude that the Witten's N=2 superalgebra may be the source of emergence of the Coulomb potential. It was proved also that the energy spectrum for the Coulomb potential in both cases may be derived by manipulations only on generators, so Dirac-Coulomb problem in both cases belong to the fully integrable theory.

**Acknowledgements:** The Author express his cordial gratitudes to professors A.Rusetski and U. Meissner for their hospitality at Bohn university, as well as to all participants of seminar there for valuable discussions and critical remarks. Moreover, I thank the National Rustaveli Foundation (Grants D1/13/02 and FR/11/24) for the financial support.